\documentclass[]{WileyASNA-v1}
\usepackage{moreverb}
\usepackage{natbib}
\usepackage{hyperref} 

\newcommand\BibTeX{{\rmfamily B\kern-.05em \textsc{i\kern-.025em b}\kern-.08em
T\kern-.1667em\lower.7ex\hbox{E}\kern-.125emX}}
\newcommand{\xmm}{{\it XMM-Newton}}
\newcommand{\chandra}{{\it Chandra}}
\newcommand{\nustar}{{\it NuSTAR}}
\newcommand{\rhoOph}{$\rho$~Oph}

\newcommand{\eltn}{Elias~29}

\articletype{Review Article}%

\received{31 August, 2018}
\revised{<day> <Month>, <year>}
\accepted{<day> <Month>, <year>}

\begin{document}

\title{The complex phenomena of YSOs revealed by their X-ray variability}

\author[1]{S. Sciortino*}

\author[1]{E. Flaccomio}

\author[1,2]{I. Pillitteri}

\author[3,1]{F. Reale}

\authormark{S. Sciortino \textsc{et al}}

\address[1]{\orgdiv{Osservatorio Astronomico di Palermo}, \orgname{INAF}, \orgaddress{Piazza del Parlamento 1, 90134 Palermo, \country{Italy}}}

\address[2]{\orgdiv{HEAD}, \orgname{Center for Astrophysics}, \orgaddress{60 Garden St, Cambridge, \state{MA}, \country{USA}}}

\address[3]{\orgdiv{Dipartimento di Chimica e Fisica}, \orgname{UNIPA}, \orgaddress{Via Archirafi 36, Palermo, \country{Italy}}}

\corres{*Salvatore Sciortino, INAF/Oss. Astr. di Palermo, Piazza del Parlammento 1, 90134, Palermo, Italy. \email{salvatore.sciortino@inaf.it}}

\abstract[Abstract]{X-ray observations of Young Stellar Objects (YSOs) have shown several complex phenomena at work.  
In recent years a few X-ray programs based on long, continuous and, sporadically, simultaneous coordinated multi-wavelengths observations 
have paved the way to our current understanding of the physical processes at work, that very likely regulates the interaction 
between the star and its circumstellar disk.  We will present and discuss some recent results based on a novel analysis of few selected 
very large flares observed with the Chandra Orion Ultradeep Pointing (COUP), on the systematic analysis of a large collection of flares 
observed with the Coordinated Synoptic Investigation of NGC 2264 (CSI 2264) as well as on the Class I/II YSO Elias 29, in the rho Oph star 
forming region, whose data have been recently gathered as part of a joint simultaneous XMM-Newton and NuSTAR large program.}


\maketitle

\section{Introduction}\label{sec1}

Rotation, magnetism and accretion produce X-ray emission
as a strong feature of young stellar objects (YSOs) yet to be fully understood. 
As a result high energy phenomena are key elements of the process of star formation because of the interplay, 
mediated by the magnetic field, between the newly born stars and their disks. 
Since the time scales of the involved phenomena are rather different, 
a proper tuned study of variability can allow to single out the many physical process at work.
This has been the focus of a series of key studies in the last decade of which I will present and discuss a few selected topical examples together with a glimpse at some recent 
ongoing studies, in a few cases part of multi-wavelength observational campaigns, made possible by data gathered with \xmm\ and \chandra\ and, very recently, with \nustar\ . 
Most of the recent advances have been possible thanks to long continuous observations of YSOs in nearby star forming regions, especially in Orion \citep{Getman+2005} and 
in \rhoOph\ \citep{Pillitteri+2010} or to long term monitoring programs as for the study of cycles (cf. \citet{Stelzer2017} and reference therein cited). In the following 
I will mostly concentrate on what we have learned and what we can still learn from the multi-wavelength studies of flares and of the Fe K$_\alpha$ 6.4 keV line. Other interesting issues, 
such as accretion and outflow processes, are discussed by \cite{Argiroffi2018}.

\section{YSO Flares}\label{sec2}

\begin{figure*}[t]
\centering
\vspace{-0.8cm}
\includegraphics[width=18cm]{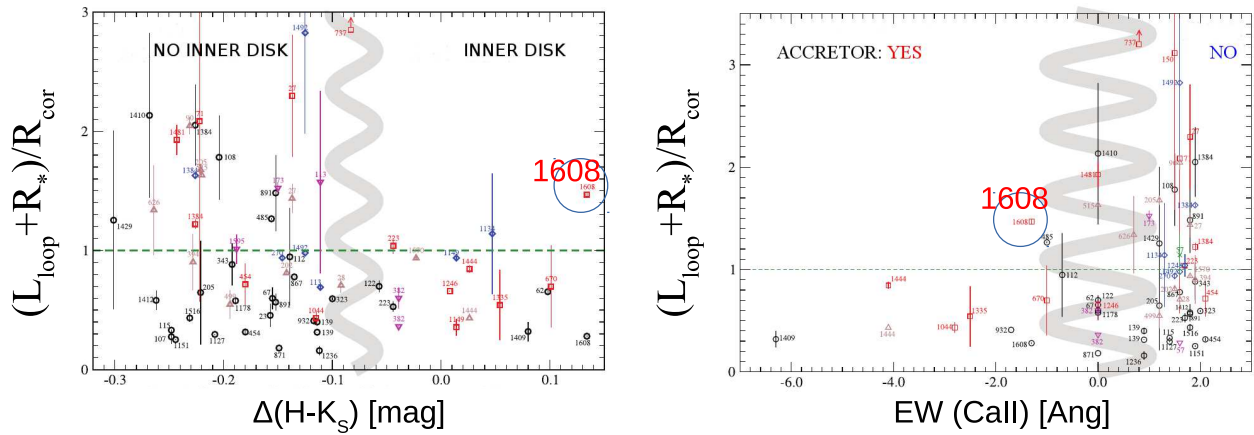}
\vspace{-1.5cm}
\caption{\label{fig1} (left) Scatter plot of $L_{loop}+R_{*})/R_{cor}$ vs. excess of ($H-K_{S}$) color, an indicator of the presence of circumstellar disk. 
(right) Scatter plot of $L_{loop}+R_{*})/R_{cor}$ vs. the EW of CaII triplet, an indicator of accretion process. 
In both panels the vertical wavy gray curve marks the transition to accreting/inner disks.
COUP 1068, that is outlined, clearly behaves very differently from the rest of (most of) 
the other sources.  COUP 332, for which the analysis derives long flaring structure, is not shown in the plots for lack of period and EW data
(adapted from \cite{Getman+2008b}.)
}
\end{figure*}
\subsection{Flares as a tool to trace the disk-star magnetosphere}

As discussed in more detail by \cite{Reale2007}, a flare is essentially an impulsive release of energy 
occurring in a tenuous plasma confined in a ``magnetic bottle'' that loses energy by optically thin radiation and by efficient thermal conduction to the chromosphere (for an extensive discussion see \citet{Reale2014}). 
The magnetic confinement is crucial for shaping the typical light curve of a flare\footnote{The shape of light curve would be very different in the case of unconfined plasma},
which is characterized by a very fast increase of the emission (due to the rapid heating of the plasma) 
followed by a slow, almost exponential decay (due to the cooling by thermal conduction and radiation).
Under very general conditions, that are typically met by most of the observed flares, 
\cite{Reale2007} showed that the time evolution of the emission and the peak temperature (and the cooling 
as traced in the $log (T) - log(\sqrt{EM})$ diagram) can provide a ``direct'' estimate of the length of 
the flaring magnetic structure (arc).
On the basis of this interpretative framework, \cite{Favata+2005} 
have analyzed a sample of 32 large flares observed on thanks to a long 
continuous \chandra\ observation toward Orion, nicknamed {\it COUP}
(PI E. Feigelson, \citet{Getman+2005}), concluding that in about 10 of them the length of the flaring structure is $3-5$ stellar radii (R$_*$). 
Similar length structures have never been ``seen'' 
in more evolved normal stars. \cite{Favata+2005} note that structures of such extent, if anchored on the stellar surface, should suffer of major stability problem due to the 
centrifugal force  since 1-2 Myr old YSOs are fast rotators with rotation period, P $\sim$ 3-6 days. Hence long loops anchored only on the star would be ripped open. 
As a solution to this problem  they conclude that, since the corotation radius of those YSOs is typically at 4-5 R$_*$, it is very likely that the loop, on which 
the flare occurs, is connecting the star and the disk (at the corotation radius). It is worth noticing that the existence of such magnetic ``funnels'' in class I-II YSOs is postulated by magnetospheric accretion scenario (e.g. \cite{Hartmann1998} and references therein cited).

A similar analysis has been performed for the several tens of YSOs of the \rhoOph \ Core F region, thanks to the data gathered with a large \xmm\ program, nicknamed {\it DROXO} (PI. S. Sciortino, 
\citet{Pillitteri+2010}). In seven YSOs \citep{Flaccomio+2009}
we have discovered intense flares. 
By means of the $log (T) - log(\sqrt{EM})$ diagram we have derived
the length of the flaring structure.
In 2 out of the 7 flares that have been studied the derived length is several stellar radii. It is worth noting that 
the fraction of the very long flaring structures in \rhoOph\ is similar to the one observed in Orion, namely $\sim$ 30\% .    

Subsequently, by adopting a new flare spectral analysis technique that avoids nonlinear parametric modeling, Getman and collaborators \citep{Getman+2008a,Getman+2008b} 
analyzed the full set of COUP data of 216 flares 
occurring in 161 YSOs  and determined the length of the flaring loop, $L_{loop}$.
Based on estimation of the stellar radius, $R_{*}$, and disk keplerian corotation radius, $R_{cor}$ 
\footnote{i.e. the distance from stellar surface at which the angular velocity of disk equates that at stellar surface}, they constructed the scatter plots of
$(L_{loop}+R_{*})/R_{cor}$ as a function of indicators of the presence of circumstellar disk or 
of on-going accretion process (cf. Fig. \ref{fig1}).
On the basis of those scatter plots they concluded that: 1) circumstellar disks have no effect on flare morphology; 2) circumstellar disks are unrelated to flare energetics;
3) super-hot (> 100 MK) ``non-standard'' flares do occur in accreting YSOs (in agreement with \citet{Favata+2005}); 4) circumstellar disks may truncate PMS magnetospheres, i.e.,
$(L_{loop}+R_{*})/R_{cor} < 1 $.
Points 1) and 4) are at odd with the findings of \cite{Favata+2005} because they seem to imply the non-existence of star-disk interconnecting flaring structures. However it is worth noticing
that even in 
the analysis of \citet{Getman+2008b} there are remarkable exceptions, namely a few YSOs whose data points cannot be reconcilied with the above conclusions, notably COUP 1688 and 
COUP 332\footnote{COUP 332 has a weak NIR counterpart and is not shown in the 
plots because of lack of period and EW(CaII) data.} (two of 10 YSOs with long flaring loop according to \citet{Favata+2005}), at this two Orion YSOs we have to add also DROXO 63 and DROXO 67, 
the two \rhoOph\ showing firm evidence of very 
long flaring loops. In summary, the issue of the existence of star-disk interconnecting flaring arch has been matter of debate over the last decade. The issue is particularly interesting since, 
depending on the actual occurrence of those large flares, they can affect the early evolution of circumstellar disks with far reaching effects even on the formation of planetary systems.

\begin{figure}[t]
\vspace{-0.3cm}
\hspace{-0.3cm}
\includegraphics[width=9cm,height=5.2cm]{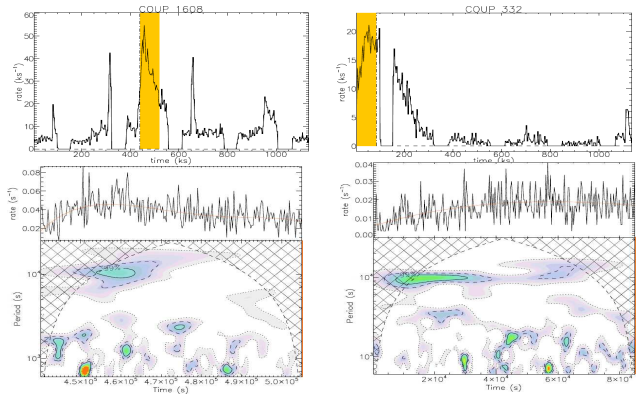}
\vspace{-0.5cm}
\caption{\label{fig2} (Left panels) The data and the summary of the wavelet analysis results for COUP 1068. Time is measured from the beginning of the observation; the analyzed flare data segment is  highlighted in yellow. The central panel shows
the analyzed data segment after subtraction of the running average, while the bottom panel shows the intensity curve as function of period and time. The 
statistical significance maps are also shown, they allow to discriminate the statistically 
significant period and the duration of the periodic signal. 
The dashed outer region is outside the so-called ``cone of influence'', delimiting the region where the analysis is meaningful. 
(Right panels) The analogous plots for COUP 332.
}
\end{figure}

Recently a novel analysis technique has allowed to further investigate the relevant astrophysical question of the existence of long flaring magnetically confined structures very likely 
interconnecting the circumstellar disk and the central star. First of all \cite{Flaccomio+2012} has performed a sophisticated time resolved analysis of all available COUP data showing that 
disk-bearing stars are definitively more X-ray variable than disk-less ones, and proposed that this can be easily explained as due to the effect of time-variable absorption by warped and rotating
circumstellar disks.
 
\begin{figure}[h]
\centering
\vspace{-1.20cm}
\includegraphics[width=7.5cm]{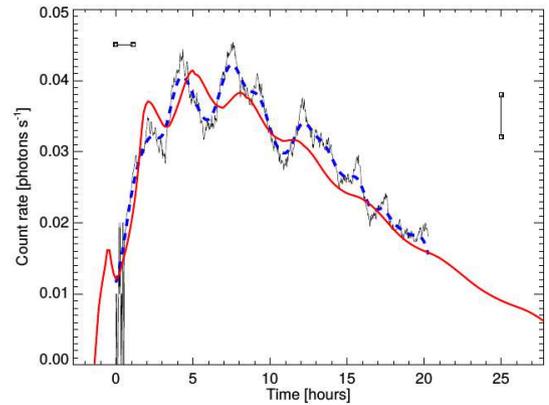}
\vspace{-0.2cm}
\caption{\label{fig3} The smoothed and running average subtracted light-curve of V772 Ori flare (black line) showing the oscillation of X-ray emission is compared with the hydrodynamical model 
synthetized light-curve in the case of a short duration heating pulse (blue dashed line) and of a long duration heating (red line) on the same long length flaring structure. 
Only a long loop with a length of about 10-20 $R_{*}$  and a short heating pulse are able to reproduce the observed oscillations both in intensity and period.}
\end{figure}

Even more relevant is a novel analysis of the light curves based on the 
so-called Morlet wavelet (eg. \cite{LopezSantiago2018} and references therein cited) of some 
of the big COUP flares studied by \citet{Favata+2005}. 
Wavelet analysis has demonstrated to be a powerful way to reveal oscillations in the
light curve of stars during coronal flares.  Indeed its application to some of the COUP big flares has shown the existence of oscillations during the flare decay phase. This has allowed 
an accurate derivation of the oscillation period from which, on the basis of simple physical argument, it is possible to derive the length of the flaring structure where the oscillating X-ray 
emission comes from \citep{Lopez-Santiago+2016}. More recently, on the basis of the wavelet analysis, \cite{Reale+2018} have reported the detection of
large-amplitude ($\sim$ 20\%), long-period ($\sim$ 3 hr) pulsations in the light curve of two day-long flares observed with COUP. Detailed hydrodynamical modeling of two flares observed on V772 Ori 
(shown in Fig. \ref{fig3})
and OW Ori shows that these pulsations track the sloshing of plasma along a single elongated 
magnetic tube, triggered by a heat pulse whose duration ($\sim1$ hr) is much shorter that the sound crossing time along the loop. From this simple and robust modeling \cite{Reale+2018} concluded that
the involved magnetic tubes are $\sim$ 20 solar radii long, and, very likely, connect the stars with their surrounding disks. 
\section{Multi-wavelength studies}

The multi-wavelength coordinated simultaneous observations are certainly an interesting territory, but poorly 
explored so far because of the factual difficulties of organizing those programs, as they  
require to access at the same time several observing facilities from space and from the ground. 
The most notably coordinated simultaneous observations that have been pursued till today are
i) CSI (Coordinated Synoptic Investigation) of NGC 2264 based on COROT, \chandra\ and Spitzer simultaneous observations (eg. \cite{Cody+2013,Cody+2014,Stauffer+2016}) of which some recent 
results, mostly based on the X-ray data, will be illustrated in the following,
ii) Kepler/\xmm \ observations of the Pleiades, where about a dozen of flares have been detected 
and studied in detail \citep{Guarcello+2018a,Guarcello+2018b},
ii) Few more selected interesting individual sources (e.g. AB Dor, V4046 Sgr, V2129 Oph, etc.), some of which are YSOs.

{\bf Systematic study of YSO flare energetics:}
\cite{Flaccomio+2018} have performed a detailed analysis of all the flares observed in the NGC~2264 YSOs during the CSI program and have compared the light curves obtained in X-rays (with \chandra), 
in optical (with COROT), and in the infrared (with Spitzer). This analysis allows deriving few conclusions, namely, i) 
the flare peak luminosity measured in the optical, IR and X-ray band-passes are tightly correlated, with a small scatter, a similar relation (with a similar amplitude scatter) 
holds also for the flare energy released in the optical, IR and 
X-ray band-passes; ii) the relationship holds over more than 3 orders of magnitude with little (.3 dex) scatter. This is somehow surprisingly given the available data and analysis assumptions; 
iii) the 
flare energy emitted in soft X-rays (i.e. in the \xmm \ bandpass) is about 10\% to 20\% of the flare energy emitted in the optical band; iv) the flare energies are up to $\sim 5$ dex higher than those of
the brightest solar flares, and the simple extrapolation of solar flares to this extreme regime requires some cautions. As an example, the data seem to indicate that the flare photospheric temperature 
is significantly lower than $10^4$ K, that is the typical solar value; v) finally there is evidence of a strong IR excesses for flares in stars with circumstellar disks: likely as a result of
the direct response (heating) of the inner disk to the optical/X-ray flare.

{\bf Unveiling circumstellar disks by time resolved X-ray spectroscopy:}
The main mechanisms responsible for the YSO X-ray variability are variable extinction, unsteady accretion, 
and rotational modulation of both hot and dark photospheric spots and X-ray-active regions. In stars with disks, this variability is related to the morphology 
of the inner circumstellar region ($\le 0.1$ AU) and that of the photosphere and corona, all impossible to be spatially resolved with present-day techniques.  Thanks to the CSI data \cite{Guarcello+2017} have studied the X-ray spectral properties during optical bursts and dips in order to unveil
the nature of these phenomena occurring on disk bearing YSOs. They have analyzed simultaneous CoRoT and Chandra/ACIS-I observations to search for coherent optical 
and X-ray flux variability. In stars with variable extinction, they have looked for a simultaneous increase of optical extinction and X-ray absorption during 
the optical dips; in stars with accretion bursts, they have searched for soft X-ray emission and increasing X-ray absorption during the bursts. 

\cite{Guarcello+2017} have found evidence for coherent optical and X-ray flux variability among the stars with variable extinction. In 38\% of the 24 stars with optical dips, 
they observe a simultaneous 
increase of X-ray absorption and optical extinction. In seven dips, it is possible to calculate the $N_{H}/A_{V}$ ratio in order to infer the composition of the obscuring material. In 25\% of the 
20 stars with optical accretion bursts, they observe increasing soft X-ray emission during the bursts arguably associated to the emission of accreting gas. 
It is not surprising that these 
properties have been observed only in a fraction of YSOs with dips and bursts, since favorable geometric configurations are required. 
The observed variable absorption during the dips is mainly due to dust-free material in accretion streams. 
In stars with accretion bursts we observe, on average, a larger soft X-ray spectral component not 
seen in non-accreting stars.

\section{Iron K{\protect$\alpha$} 6.4 keV Fluorescence Line}

Fe K$\alpha$ 6.4 keV line from ``cold'' material (with EW$>$ 100 eV) has been found in tens of YSOs, mostly in Orion and \rhoOph \, but the relation between 
line with EW $>$ 100 eV and flare is quite controversial: 
in YLW16A in \rhoOph\ the line has been seen during an intense X-ray flare \citep{Imanishi+2001};
in 7 YSOs in Orion the line has been seen during flares \citep{Tsujimoto+2005};
in Elias 29 in \rhoOph\ the line has been seen during quiescence and flaring periods \citep{Favata+2005b,Giardino+2007}, 
the same is true for many other YSOs in Orion \citep{Czesla+Schmitt2010}.

If photoionized, then the EW is $\leq$ 100 eV in the case of a corona exciting ``photospheric material'' 
\citep{Drake+2008} and the EW is $\leq$ 150 eV for an AGN disk excited by a power law source
\citealp{Matt+1991, George+Fabian1991}. 
A Fe K$\alpha$ line with with EW $>$ 100 eV has never been seen in ``normal'' stars 
while the Fe K$\alpha$ line with EW below 100 eV has been seen in a few ``active'' stars.

We have to face the unsolved question of ``how can the Fe K$\alpha$ EW be > 100 eV ? and even > 800 eV ?''. 
As matter of fact the current data leave open the scenarios where photo-ionization alone could be 
insufficient to explain such a strong fluorescent emission and collisional excitation is required. 
\citet{Drake+2008} have considered the Fe K$\alpha$ fluorescent line emission in the relatively few cases known at the time concluding that there was not
compelling evidence for a collisional excited fluorescence from high energy electrons. They have proposed 4 different possible 
explanations, namely: 1) Super-solar Fe abundance in disk material, but an extremely high abundance of Fe is required  and the EW rapidly saturates at $\sim$ 800 eV \citep{Ballantyne+2002};
2) Disk Flaring where a favorable geometry results in a source solid angle > 2 $\pi$, but this can increase line intensity by, at most, a factor two or three;
3) Line emission due to an "unseen" flare obscured by stellar disk. This implies that the evaluation of the exciting continuum is 
grossly underestimated, but a very ad-hoc geometry is required;
4) Excitation due to high energy non-thermal electrons, but this requires a substantial amount of 
energy stored in the impinging particles \citep{Ballanyne+Fabian2003}.

It is worth to notice that i) only solution n. 4) may, in principle, explain the EW of $\sim$ 1400 eV found in V1486 Ori \citep{Czesla+Schmitt2007} and ii) that solutions n. 2 and 3, requiring ``ad-hoc'' geometries,
are unsatisfactory when the fluorescent emission becomes quite common, as the accumulated data clearly indicate.

\subsection{Elias 29: DROXO main results}

One of the most intriguing results obtained with the DROXO, 500 ks long, observations, of the \rhoOph\ core F, is the fact that in Elias~29, 
a class I/II YSO that is seen almost face-on, we have found period of ``quiescent'' as well as of ``flaring'' emission and have found that the \xmm\ spectra requires, to be 
adequately fitted, the presence of the Fe K$\alpha$ line at (about) 6.4 keV. The line equivalent width, EW, does vary with time, as time resolved spectroscopy with a ``resolution'' of about 60 
ksec has clearly shown. Since Elias~29 is the strongest YSO showing this phenomenon it is clearly a key target for any further investigation.

\subsection{Elias 29: the new \xmm\ and \nustar\ joint program}

\begin{figure}[h]
\centering
\vspace{-0.5cm}
 \includegraphics[width=7.cm]{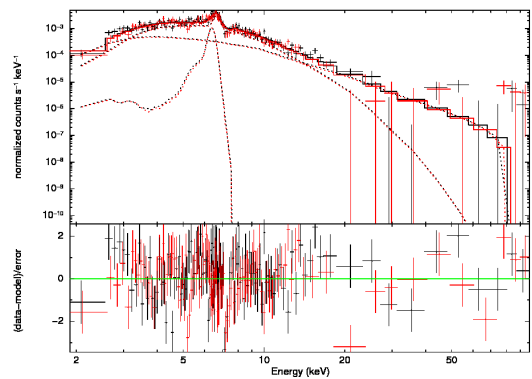}
\vspace{-0.2cm}
 \caption {\label{fig4} \nustar\ FPM A and B spectra of Elias~29 with the best fit model. 
The model is composed by an absorbed 1T APEC thermal component plus a  Gaussian line at 6.4 keV 
and a power law in order to model also the region above 20 keV. 
While the emission up to $\sim$ 20 kev is well described from the derived \xmm\ best fit model, 
it is clear the presence of a statistically significant ``non-thermal'' excess above 20 keV.} 
\end{figure}

Before the availability of \xmm\ and \nustar\ simultaneous observations to test for the presence of a non thermal population of electrons responsible for the excess of fluorescence of a disk-bearing YSO was not possible. 
\nustar\ has offered for the first time the opportunity to perform this investigation.
In this context we have obtained joint and simultaneous \xmm+\nustar\ 300 ksec long observations of \eltn\ 
devoted to acquiring spectra from soft (\xmm\ band 0.3-8.0 keV) to hard (\nustar\ band 3-80 keV) X-rays. 
Our main aims were to detect any non thermal hard X-ray emission from \eltn, to study any time 
variability that could occur and to relate
these features to the fluorescent emission with the aim to explain its origin. 
The interested reader will found all the details of the analysis performed and a detailed account of results in \cite{Pillitteri+2018}. Here we will just provide a summary of
the relevant findings. We have found evidence of an excess of likely ``non-thermal'' emission 
above 20 keV in the \nustar\ spectra (Figure \ref{fig4}). The presence of the excess 
does not seem to be associated to the occurrence of flares and we confirm the presence of a line emission at about 6.4 keV whose EW does vary with time and that does not peak 
at the maximum flare intensity.

We have also investigated if the available \xmm\ spectra do allow 
to trace possible time variation of the centroid of the Fe fluorescent line. 
We have performed an extensive sets of simulations and have explored a range of accumulated counts and of EWs 
of the fluorescent line. The input spectrum is the best fit model spectrum of Elias~29 plus a Gaussian line with a centroid at 6.4 keV. The simulations
shows that, if the source spectrum contains, in the 5-8 keV range, more than 500 counts and if the EW $>$ 300 eV then it is extremely unlikely that the fitted line centroid is above 6.5 keV.
Since the fitted line centroids are above 6.5 keV in a number of data segments with more than 500 counts we conclude that there is a convincing evidence that a non-negligible fraction of the material
emitting the fluorescent line is not in a neutral state. As calculated and discussed by 
\cite{Kallman+2004} a centroid at $\sim$ 6.5 keV would imply, if ionization equilibrium condition are 
met, that emitting Fe is at 10$^5$ K.

\section{Conclusions}

Over the last decade our knowledge of high energy phenomena occurring in YSOs has greatly advanced. From the many painstaking efforts we have learned some general lessons 
and from observational efforts we have shed light on some controversial issues while others still remain unsolved and will require further 
investigations. Limiting ourselves to the issues we have discussed we can conclude that:
1) Long (> 100 ks) continuous observations catch many kind of variability at work in YSOs and allow us to unveil the nature of the physical processes behind them.
2) Novel studies of big flares in YSOs firmly confirm that very elongated (arch-like) structures exist and are involved in those flares.
Due to simple considerations on the effect of centrifugal force it is likely that those structures connect the star and the disk at the co-rotation radius.
3) Simultaneous coordinates observations have shown to be crucial to improve our understanding of the nature of YSO flares and their effects on disk evolution.
4) Strong and variable Fe K$\alpha$ line is a common feature of disk-bearing YSOs. Both the nature of the excitation mechanism and the physical state of emitting matter 
remain far from being clear. There is a growing evidence that the ionization stage of the emitting gas is different from being mainly neutral.

We are confident that in the next decade the ATHENA observatory 
\citep{Barcons+2017} with its high throughput and X-IFU high spectral resolution 
\citep{Barret+2016}  will allow to answer all 
those and many more questions in the field of star formation and evolution \citep{Sciortino+2013}.

\section{Acknowledgments}

We acknowledge modest financial contribution from the agreement ASI-INAF n.2017-14.H.O. IP acknowledges support from the ASI and the Ariel Consortium.

\bibliography{SSciortino}

\end{document}